\documentclass[12p]{article}
\usepackage{amssymb}
\usepackage{amsmath}
\usepackage{amsthm}
\usepackage{graphicx}

\newtheorem{theorem}{Theorem}

\newtheorem{corollary}{Corollary}
\newtheorem{assumption}{Assumption}

\newtheorem{definition}{Definition}
\usepackage{color}
\usepackage{lineno}

\usepackage{times}
\usepackage{bm}
\usepackage{natbib}
\usepackage{ulem}
\usepackage[plain,noend]{algorithm2e}

\makeatletter
\renewcommand{\algocf@captiontext}[2]{#1\algocf@typo. \AlCapFnt{}#2} 
\def\@algocf@capt@plain{top}
\renewcommand{\algocf@makecaption}[2]{%
  \addtolength{\hsize}{\algomargin}%
  \sbox\@tempboxa{\algocf@captiontext{#1}{#2}}%
  \ifdim\wd\@tempboxa >\hsize
    \hskip .5\algomargin%
    \parbox[t]{\hsize}{\algocf@captiontext{#1}{#2}}
  \else%
    \global\@minipagefalse%
    \hbox to\hsize{\box\@tempboxa}
  \fi%
  \addtolength{\hsize}{-\algomargin}%
}
\makeatother

\def\Pr{\textnormal{pr}}

\def\mc{\mathcal}
\def\bs{\boldsymbol}
\def\mb{\mathbb}

\def\kk{_k}
\def\kl{_{k+1}}

\def\ki{_{k,i}}
\def\KK{_K}
\def\Ki{_{K,i}}
\def\one{_1}

\def\V{\mc V} 
\def\E{\mb E} 

\def\argmax{\arg\max}

\def\CatRisk{c_1} 
\def\CNK{c_2} 
\def\nNK{N} 

\usepackage{url} 

\usepackage{xr}
\begin{document}



\markboth{H. Cho, S. T. Holloway, D. J. Couper, \and M. R. Kosorok}{Optimal dynamic treatment regimes for survival outcomes with dependent censoring}

\title{Multi-stage optimal dynamic treatment regimes for  survival outcomes with dependent censoring}

\author{Hunyong Cho\thanks{\scriptsize{Department of Biostatistics, University of North Carolina at Chapel Hill,
    hunycho@live.unc.edu}},
    Shannon T. Holloway\thanks{\scriptsize{Department of Statistics, North Carolina State University}},
    David J. Couper\thanks{\scriptsize{Collaborative Studies Coordinating Center, Department of Biostatistics, University of North Carolina at Chapel Hill,
    david\_couper@live.unc.edu}},\\
    ~ \\
    and Michael R. Kosorok\thanks{\scriptsize{
    Department of Biostatistics,
    Department of Statistics and Operations Research,
	University of North Carolina at Chapel Hill}}}





\maketitle

\begin{abstract}
We propose a reinforcement learning method for estimating an optimal dynamic treatment regime for survival outcomes with dependent censoring. 
The estimator allows the failure time to be conditionally independent of censoring and dependent on the treatment decision times, supports a flexible number of treatment arms and treatment stages, and can maximize either the mean survival time or the survival probability at a certain time point. The estimator is constructed using generalized random survival forests and can have polynomial rates of convergence. Simulations and data analysis results suggest that the new estimator brings higher expected outcomes than existing methods in various settings. An R package \texttt{dtrSurv} is available on CRAN.
\end{abstract}

\noindent%


\section{Introduction}
Multi-stage treatments are becoming more prevalent in medicine. 
Chronic conditions such as cancer, auto-immune disease, HIV, and heart disease, often require multiple treatments over a long period. 
Cancer patients may receive multiple rounds of chemo- and/or biological therapies \citep{habermann2006, huang2020}. Auto-immune disease patients might receive additional treatments in response to relapses \citep{edwards2006, hogan2013}. Traditionally such chronic diseases have been treated following a one-size-fits-most approach by which the standard of care is selected as the treatment or treatment plan that will likely benefit most patients with a similar condition, i.e., the treatment that is optimal only in the sense of the marginal effects.

However in recent years, a more personalized approach to the treatment of chronic diseases has become of interest. 
Dynamic treatment regimes formalize a data-driven approach to treatment by optimizing the overall outcome of interest and providing treatment suggestions that are based on the patient's information available at each point of the decision making process \citep{murphy2003, kosorok2015, kosorok2019, tsiatis2019}. In part because they utilize all available information, such optimized dynamic treatment regimes are often found to be more beneficial than a traditional set of fixed treatment plans \citep{kidwell2015}. Further, because optimal dynamic treatment regimes focus on the overall or long-term outcome, they are more advantageous than applying multiple single-stage treatment rules that only optimize the stage-level outcomes.


Finding an optimal dynamic treatment regime is a reinforcement learning problem \citep{sutton2018}, wherein an action by an agent, or a physician, changes both the immediate reward and the environment of the next stage and the learner, or the statistician, searches for the rule of actions that brings the most overall reward. 
Reinforcement learning problems are often solved using the Q-learning algorithm \citep{watkins1989, zhu2015, zhao2011, zhang2017, qian2011}, which defines the value as the expected cumulative sum of discounted rewards and optimizes the set of rules using backward recursion.

There are some challenges in using Q-learning in biomedical research that involves survival outcomes.
As the failure time is often censored, any approach should take into account the missing information, for example, using redistribution of the failure probability as in the Kaplan-Meier \citep{efron1967, robertson1984} or inverse probability weighting \citep{robins1994, robins1992, wahed2006, orellana2010}. Furthermore, in multi-stage contexts, patients may not be able to make all the  planned visits due to failure or dropping out of the study, making the total number of treatments received to vary among patients. Without addressing these missing information issues, backward recursion in Q-learning is not readily applicable.

Another significant challenge posed in survival analysis is that the treatment timing may not be fixed, may even depend on a patient's status or previous treatment history, and could have association with the failure time. In such scenarios, the three events, failure, proceeding to the next treatment, and censoring, may have mutual associations, and they compete so that only the earliest event is observed. For example, a cancer patient whose physical condition has improved has a longer expected survival time and, at the same time, may want to take another round of chemotherapy earlier than expected.
However, the full joint distribution of those events is not identifiable without knowledge of their dependence structure \citep{tsiatis1975}. Thus, the dependency between the events needs to be carefully addressed.

Several methods regarding the estimation of dynamic treatment regimes for survival outcomes have been proposed in the literature. \cite{goldberg2012} first developed a dynamic treatment regime  estimator for survival outcomes using a Q-learning framework. To address the problems introduced by a different number of treatments as well as by censoring, 
the authors proposed modifying the survival data so that each observation has the same number of treatment stages without missing values and using a standard Q-learning framework along with inverse probability weighting to account for censoring.
However, censoring was assumed to be completely independent of all covariates and event times. The estimated Q-function and the resulting decision rules would be thus subject to bias for data with dependent censoring.

\citet{huang2014} addressed a similar problem using backward recursion. Their motivating problem was a recurrent disease clinical trial where patients receive an initial treatment followed by a salvage treatment if patients experience either treatment resistance or relapse. 
\citet{simoneau2019} extended the dynamic treatment regime estimator for non-censored outcomes via weighted least squares \citep{wallace2015} into the censored time-to-event data setting. Both of these methods use a linear accelerated failure time model, which carries a risk of model misspecification and results in a restrictive policy class. Thus, a more flexible model is needed to allow for more relaxed distributional assumptions.
\cite{wahed2013} proposed a dynamic treatment regime estimator using full specification of the likelihood and \citet{xu2016} developed a Bayesian alternative, where the disease progression dynamic was modeled using a dependent Dirichlet process prior with a Gaussian process measure. However, their policy classes are finite-dimensional and, thus, may not be able to consider diverse patient characteristics and contexts sufficiently. 

Each of the existing methods described above has its limitations. First, not all methods allow for a flexible number of treatment stages and/or treatment levels. For example, the methods in \citet{wahed2013} and \citet{huang2014} were designed for a two-stage decision problem, and the \citet{simoneau2019} approach is restricted to binary treatments.
Second, the outcomes of interest in most methods are limited to the mean survival time. \citet{jiang2017} proposed an optimal dynamic treatment regime estimator that maximizes the $t$-time survival probability. However, none of the methods permit choosing the criterion except \citet{wahed2013}.
Third, strong structural assumptions are employed for all of these methods, with the exception of \citet{goldberg2012}. Assumptions, such as the accelerated failure time model or the proportional hazards models used by the existing methods, are subject to model-misspecification, may limit the policy class, and, consequently, may invoke loss in the value of the optimal policy.
Finally, the censoring assumption of \cite{goldberg2012} is restrictive. Censoring is associated with failure time in many medical applications, although it is often reasonable to assume independence of censoring after conditioning on patient historical information. See Table \ref{tab:methods} for a summary of those methods and assumptions.

\begin{table}
\def~{\hphantom{0}}
{\scriptsize
    \begin{tabular}{lcccccc}
    \hline
    method & $K$ & $|\mc A\kk|$ & criterion & $T$ & $C$ & $\{\bs\pi\}$\\
    \hline
    the new method & finite& finite& $\E[T\wedge \tau], S(t_0)$ &  NP & CI & flexible\\ 
    \citet{goldberg2012} & finite & finite & $\E[T\wedge \tau]$ & NP & independent& flexible\\
    \citet{huang2014} & 2+ & 2+ & $\E[T]$& AFT & CI & linear\\ 
    \citet{jiang2017} & finite & 2 & $S(t)$ & PH & CI & linear\\ 
    \citet{simoneau2019} & finite & 2 & $\E[T]$ & AFT& CI & linear\\ 
    \citet{wahed2013} & finite & finite & $\E[T]$, $S(t_0)$ & AFT & CI & fixed\\ 
    \citet{xu2016} & finite & finite & $\E[T]$& BNP & CI & fixed\\ 
    \hline
    \end{tabular}
  }
\label{tab:methods}
\caption{Assumptions of the existing and the proposed methods. 
$K$, the number of stages; $|\mc A\kk|$, the number of treatment arms at stage $k$; criterion, the target value to be optimized; $T$, failure time model; $C$, censoring assumption; $\{\bs\pi\}$, policy class;
2+, can be generalized to more than two;
NP, non-parametric; AFT, accelerated failure time; PH, proportional hazards; BNP, Bayesian non-parametric; 
CI, conditional independence; fixed, a fixed number of distinct treatment rules.
}
\end{table}

In this manuscript, we develop a general dynamic treatment regime estimator for censored time-to-failure outcomes that addresses all of the limitations discussed for the existing estimators. Our estimator maximizes either the mean survival time or the survival probability at a certain time using backward recursion. As the objective may not be expressed as the sum of intermediate rewards, a standard Q-learning algorithm is not applicable. Instead, the conditional survival probability information, rather than the summarized Q-values, is appended to the previous stage information. A generalized random survival forest is developed for this task, where survival curves for each individual, instead of just observed survival or censoring time, is fed into the random survival forest. A general implementation of our method is available on CRAN, R package \texttt{dtrSurv}.

The key contributions of our work are the flexibility of the proposed method and the exposition of its theoretical properties. 
The target value can be either the mean survival time or the survival probability at a certain time. The method allows for a flexible number of treatment stages as well as a flexible number of treatment arms at each stage. Censoring can be conditionally independent. The conditional survival probability is modeled nonparametrically using a random survival forest-based algorithm. 
We further show the consistency of the random survival forests and the estimated regime values as well as their rates of convergence.

The remainder of the manuscript is organized as follows. In Section~\ref{s:method}, we give notation and describe the dynamic treatment rule estimators. Theoretical properties of the estimators are derived in Section~\ref{s:theory}, and their performance is illustrated through simulations in Section~\ref{s:sim}. Section~\ref{s:examples} is devoted to the application of the proposed method to a cardiovascular disease cohort study and a two-stage leukemia clinical trial. We conclude the paper by discussing future research directions in Section~\ref{s:discussion}.

\section{The method}
\label{s:method}
\subsection{Data setup and notation}
\label{ss:notation}
We assume that each of $n$ independent patients can have a maximum of $K$ visits, or treatment stages, with a study length $\tau > 0$. At the $k$th stage, $k = 1,2,...,K$, the $i$th patient, among the $n\kk\le n$ available patients, receives treatment $A\ki \in \mc A\kk$, if he or she has survived and has not dropped out by the beginning of the stage, where $\mc A\kk$ is the finite treatment space for the $k$th stage. Throughout this manuscript, we often drop the subject index $i$ when it does not cause confusion. Our interest lies in  estimating the survival distribution $S^{\bs\pi}$ of the overall failure time $T$ of patients if they followed a dynamic treatment regime $\bs\pi = (\pi\one, \pi_{2}, ... \pi\KK)^\top$ and then finding the `best' rule $\bs\pi^*$, which optimizes a certain criterion $\phi$. An (optimal) dynamic treatment regime is a series of stage-wise policies $\pi_k$ that map historical information $\bs h_k \in \mc H_k$ to a (best) treatment $a_k\in \mc A_k$.
For $\phi$, we consider $\phi^\mu(S) = -\int_{t>0} (t\wedge \tau) dS(t)$ and $\phi^{\sigma,t_0}(S) = S(t_0)$ for some $0<t_0\le\tau$.
Without loss of generality, we assume that a prolonged time to event is preferred; thus the objective is to maximize $\phi(S)$ over the regimes.

We observe the time to next treatment $U\ki$ of the $i$th patient which is a random or fixed quantity that depends on the historical information $\bs H\ki\in \mc H\kk$  available at the beginning of the $k$th stage and the treatment assignment $A\ki$. 
At the $k$th stage, we observe either the $i$th patient's failure, advancement to next treatment, or censoring. The times from the beginning of the stage to each of these events are denoted as $T\ki$, $U\ki$, and $C\ki$, respectively. The length of the stage $X\ki$ for the $i$th patient is defined by the minimum of these times, and the hypothetically uncensored stage length is defined as $V\ki = T\ki \wedge U\ki$. 
For the last stage, $V\Ki = T\Ki$. We denote the $k$th stage censoring and treatment indicators as $\delta\ki = 1(V\ki \le C\ki)$ and $\gamma\ki = 1(T\ki \le U\ki)$, respectively. When $C\ki < V\ki$, $\gamma\ki$ is not observable. The baseline time at stage $k$ is defined as $B\ki = \sum_{j=1}^{k-1}X_{j,i},  k>1$, and $B_{1,i}=0$. 
Let $T_i,$ $C_i$,
$X_i = T_i \wedge C_i,$ and $\delta_i = 1(T_i \le C_i)$ 
denote the overall failure time, overall censoring time, overall observed time, and overall censoring indicator, respectively. We denote $\bs Z\ki \in \mc Z\kk$ as the covariate information of the $i$th patient that is newly available at the beginning of the $k$th stage. Thus, $\bs H\ki$ may include historical information such as $A_{k',i}$, $B_{k',i}$, $\bs Z_{k',i}$, $k'=1,2,...,k-1$ and $\bs Z\ki$; where $d\kk$ denotes the dimension of $\bs H\ki$ for $k= 1,2,..,K$. The number $n\kk$ of patients eligible for treatment assignment at the beginning of stage $k$ is $\sum_{i=1}^n 1(U\ki \le T\ki \wedge C\ki)$. 
See Figure~\ref{fig:diagram} for an illustration. The notation is summarized in Table~\ref{stab:notation} of the Supplementary Material.
\begin{figure}
    \centering
        \includegraphics[width = 0.9\textwidth]{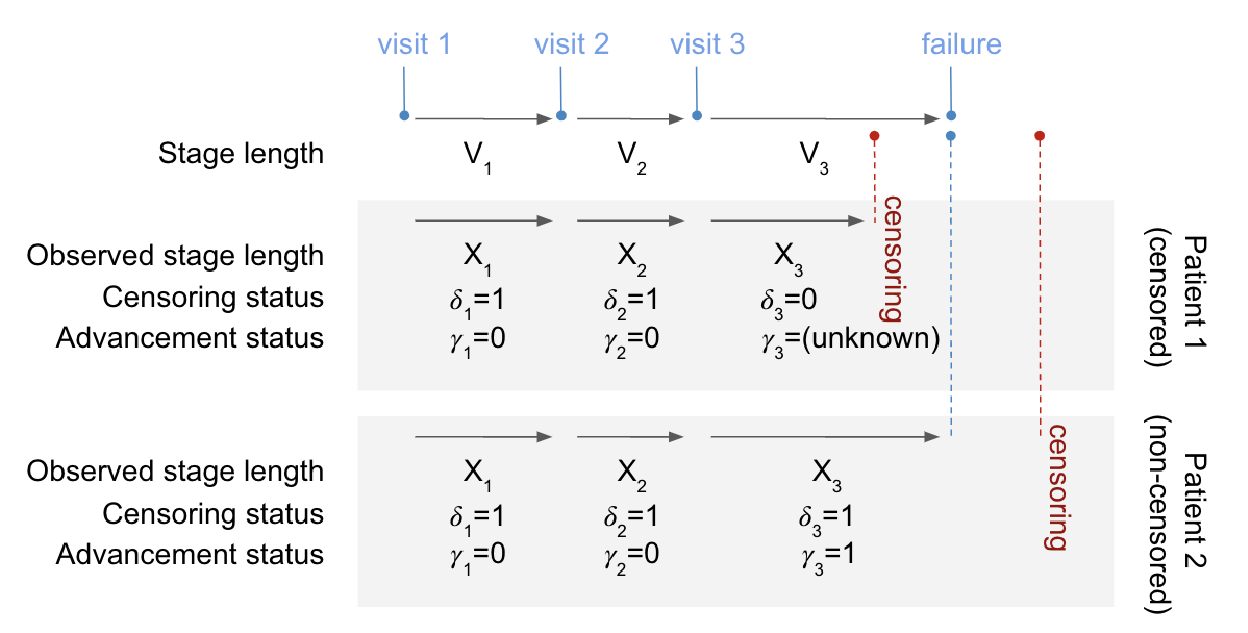}
    \caption{An examples of two patients with different censoring statuses after their third visit.}
    \label{fig:diagram}
\end{figure}

\subsection{Overview of the method}
\label{ss:method_overview}
As is the case in Q-learning, our estimator optimizes the intermediate stage decision assuming that all the later decisions are optimal, and thus backward recursion is used. 
Our method is based on the observation that the remaining life
$L\kk$ at stage $k$ is defined recursively as a convolution of $L\kl$ and the uncensored current stage length $V\kk$ and that
\begin{align}
S\kk(t\mid \bs H\kk, A\kk, \delta\kk = 1)
  =& \int S\kl(t - V\kk\mid \bs H\kl, A\kl, \delta\kk = 1)\nonumber\\
  & \hspace{1cm} dP(V\kk, \bs H\kl, A\kl \mid  \bs H\kk, A\kk, \delta\kk = 1), \label{eq:marginal_nocensor} 
\end{align}
where $S\kk (t\mid \cdot) = 1$ for all $t<0$, and $S\kk (t\mid \gamma_{k-1} = 1) = 1(t <0)$, for $k=K-1, K-2, ..., 1.$ 
Under the covariate-conditionally independent censoring assumption, the contribution of the censored cases is incorporated into the equation by redistributing the probability to the right \citep{efron1967}:
\begin{align}
S\kk(t\mid \bs H\kk, A\kk)
   = & \int \bigg[\delta\kk S\kl(t - X\kk\mid \bs H\kl, A\kl)  \nonumber\\
    & ~ ~ ~ ~ ~ + 
   (1-\delta\kk)\Big\{1(X\kk > t) + 1(X\kk\le t) \frac{S\kl(t\mid \bs H\kl, A\kl)}{S\kl(X\kk\mid \bs H\kl, A\kl)}\Big\} \bigg]\nonumber\\
   &~ ~ ~ \times dP(X\kk, \bs H\kl, A\kl, \delta\kk \mid  \bs H\kk, A\kk). \label{eq:marginal} 
\end{align}
This is the key formula we rely on in optimizing the distribution of $L_k$. This convolution makes it unnecessary to estimate the joint distribution of $U_k$ and $T_k$. Throughout this manuscript we assume that $L_k$ is independent of $C_k$ conditionally on $\bs H_k$ and $A_k$, $k=1,2,...,K$.

Assuming the standard causal inference assumptions, which we will formally introduce in Subsection~\ref{ss:consist2} (Assumption \ref{a:sutva}--\ref{a:positivity}), we can translate $S\kk(t\mid \bs H\kk, A\kk = a)$ into the counterfactual quantity $S\kk^a(t\mid \bs H\kk)$.
Thus, the regime is estimated using the following procedure: 1. estimating the conditional survival distribution 
$S\kk(t \mid \bs H\kk, A\kk)$ of the remaining life $L\kk$ for $k=K$; 2. optimizing the survival distribution over $\mc A\kk$ to have $\hat \pi\kk$ and $\hat S\kk^*(t \mid \bs H\kk) \equiv \hat S\kk(t \mid \bs H\kk, A_k = \hat \pi\kk(\bs H\kk))$ for $k=K$; 3. augmenting the estimated optimal curves by the previous stage lengths, i.e., $\hat S\kk^*(t - X_{k-1} \mid \bs H\kk, A\kk)$ for $k=K$; and, 4. repeating steps 1--3 for $k=K-1,K-2,...,1$. A high-level summary is provided in Algorithm~\ref{al:dtrSurv}, and we describe the regime estimation procedure more in detail next.

\begin{algorithm}
    \label{al:dtrSurv}
    \SetAlgoLined
    \KwResult{A dynamic treatment regime estimate $\hat{\bs\pi} = (\hat\pi_{1},\hat\pi_{2},...,\hat\pi_{K})^\top$}
     \For{ stage $k = K, K-1, ..., 1$}{
      Input: $\Big\{\Big(\bs H\kk, A\kk, \delta\kk, (\hat S\kl^*(\cdot - X_{k} \mid  \bs H\kl)\Big)\Big\}_{i=1}^{n\kk}$ for $k < K$ and $\{(\bs H\kk, A\kk, \delta\kk, V\kk)\}_{i=1}^{n\kk}$ for $k=K$} \;
      Obtain $\hat S\kk(\cdot \mid \bs H\kk, A\kk)$ via generalized random survival forest (Algorithm S\ref{al:grsf})\;
      Obtain $\hat \pi\kk(\bs h) = \argmax_{a\in\mc A\kk} \phi\kk\Big\{\hat S\kk(\cdot\mid \bs H\kk = \bs h, A\kk = a) \Big\}$\;
      Define $\hat S\kk^*(\cdot \mid \bs H\kk) = \hat S\kk(\cdot \mid \bs H\kk, A\kk= \hat \pi\kk(\bs H\kk) )$\;
     
     \caption{the proposed dynamic treatment regime estimator.}
\end{algorithm}

At the beginning of the last stage, there are $n\KK$ patients who have neither experienced failure nor been censored. The data for these patients are used to estimate the terminal stage covariate-conditional survival probability $S\KK(\cdot\mid \bs H\KK, A\KK)$ for each treatment arm using random survival forests. 
Then the optimal regime for the final stage is estimated by $\hat\pi\KK = \argmax_{a \in \mc A_K} \phi\KK\Big\{\hat S\KK(\cdot, a)\Big\},$ where $\hat S\kk$ is the estimate of $S\kk$ and $\phi\kk$ is defined adaptively for the intermediate and terminal survival probabilities, $k=1,..,K$. Specifically, $\phi^\mu\kk\Big\{S\kk(\cdot \mid \bs H\kk)\Big\} = B\kk - \int_{t>X\kk}(t\wedge\tau)dS\kk(t\mid \bs H\kk)$ and $\phi^{\sigma, t_0}\kk\Big\{S\kk(\cdot\mid \bs H\kk)\Big\} =S\kk(t_0 - B\kk \mid \bs H\kk)$. In Proposition \ref{as:prop1} of the Supplementary Material, we show that backward recursion is legitimate for the choices of $\phi = \phi^\mu, \phi^{\sigma, t_0}$.

The optimized survival information of the final stage, $\hat S\KK^*$, is carried back to the previous stage; the $i$th patient 
with a stage length $X_{K-1,i}$
is given a probabilistic augmentation so that the resulting survival probability of remaining life $L_{K-1}^{\hat \pi_K}$ at the second to the last stage is $\hat S_{K}^*(t - X_{K-1,i}\mid \bs H\Ki)$. The probabilistic augmentation can be thought of drawing single ($B=1$) or many ($B>1$) $L_{K-1,i,(b)}$'s from $\hat S_{K}^*(t -X_{K-1,i}\mid \bs H\Ki)$ for subject $i$ so that $\{L_{K-1, i, (b)}: b=1,2,...,B\}$ is obtained with an equal weight $\frac 1 B$. 
For those who have already experienced failure or censoring by the end of the stage, no augmentation is required.

Next, the conditional distribution of the remaining life from the beginning of the $(K-1)$st stage is estimated by the random survival forests based on the $n_{K-1}$ patients. Then estimation of $S\kk, \pi\kk, $ and $S\kk^*$ is recursively performed until we reach the first stage and obtain $\hat {\bs \pi} = (\hat \pi_{1}, \hat \pi_{2}, ... , \hat \pi_{K})^\top$.

For the intermediate stages ($k<K$), however, unlike the last stage, the patients' remaining life $L_{k}$ is not observed as a single value but is given in the form of a stochastic process, or a set of random draws per person. For this specific type of data, we develop a generalized random survival forest that implements this analytically without random draws. See Section \ref{as:grsfmethod} of the Supplementary Material for further details. We also note that having single draws $L_{K,i,(b)}$ per person, in fact, is sufficient to derive the rate of convergence, and the random forest estimation $\hat S_{k,(b)}$ that uses single draws could be easily done using off-the-shelf software. However, since estimation based on multiple draws $\hat S_{k}$ is usually favorable in terms of precision, we use $\hat S_{k}$ for data analyses.

One consideration for the $\phi^{\sigma,t_0}$ criterion is that the choice of $\phi$ may not be practical for patients who already survived $t_0$, since treatment decision becomes irrelevant for them with such a criterion. Thus a composite criterion, $\phi^{\sigma, t_0, \mu}(S) = \{S(t_0), \E[T\wedge \tau]\}$, of firstly maximizing $S(t_0)$ until time $t_0$ and secondly maximizing $\E[T\wedge \tau]$ afterward is suggested in practice, where the values are evaluated lexicographically.

\section{Theoretical properties}
\label{s:theory}
We show consistency and rates of convergence of the generalized random survival forests and the proposed dynamic treatment regime estimator $\hat {\bs \pi}$. The consistency of a regime is established  in terms of its value, which will be defined in Subsection \ref{ss:consist2}. 
As consistency and the rate of convergence are developed using different theoretical approaches, we give two different sets of assumptions.

\subsection{Convergence properties of generalized random survival forests}
\label{ss:consist1}

We have assumptions regarding the tree-based estimator terminal node size and splitting rules. The first assumption forces the terminal node size to be both sufficiently large absolutely and sufficiently small relative to the sample size.

\begin{assumption}[Terminal node size]
\label{a:terminalnode}
The minimum size $n_{\text{min}}$ of the terminal nodes grows at the following rate:
$n_{\text{min}} \asymp n^\beta, \frac 1 2 < \beta < 1,$
where $a \asymp b$ implies both $a = \mc O(b)$ and $b = \mc O(a)$.
\end{assumption}

Next, we give definitions of regular trees and random split trees that are often used for consistency proofs in the random forest literature (\cite{meinshausen2006}, \cite{wager2015}, \cite{cui2017}, \cite{wager2018}).

\begin{definition}[Random-split and $\alpha$-regular trees and forests]
\label{def:random}
A tree is called a \textit{random-split} tree if each feature is given a minimum probability ($\varphi/d$) of being the split variable at each intermediate node, where $0<\varphi<1$ and $d$ is the dimension of the feature space. A tree is \textit{$\alpha$-regular}, if every daughter node has at least $\alpha$ fraction of the training sample in the corresponding parent node. A random forest is called a \textit{random-split}  (\textit{$\alpha$-regular}) forest, if each of its member trees is \textit{random-split}  (\textit{$\alpha$-regular}).
\end{definition}

\begin{assumption}[$\alpha$-regular and random split trees]
\label{a:split}
Trees are $\alpha$-regular and random-split with a constant $0<\varphi<1$.
\end{assumption}




\begin{assumption}[Lipschitz continuous and bounded survival and censoring probability]
\label{a:lipschitz2}
There exist constants $L_S$ and $L_G$ such that
$|S\kk(t\mid \bs h_1)-S\kk(t\mid \bs h_2)|\le L_S \|\bs h_1-\bs h_2\|$ and $|G\kk(t\mid \bs h_1)-G\kk(t\mid \bs h_2)|\le L_G \|\bs h_1-\bs h_2\|$ for all $\bs h_1, \bs h_2\in \mc H\kk$, $t\in[0,\tau - B\kk]$, and $k=1,2,...,K$, where $G\kk$ is the censoring survival distribution at the $k$th stage, and  $S\kk(\tau^- - B\kk| \bs h_k)G\kk(\tau^- - B\kk| \bs h_k) > \CatRisk$ for all $\bs h_k$ with $\gamma_{k-1}=0$---i.e., patients who made visit $k$---and some constant $\CatRisk > 0$.
\end{assumption}

\begin{assumption}[Weakly dependent historical information at each stage]
\label{a:covariate}
The patient history information $\bs H\kk$ at stage $k$ is given as a $d\kk$-dimensional vector lying in a subset $\mc H\kk$ of $[0,1]^{d\kk}$, 
and its joint ($f_{\bs H\kk}$) and marginal ($f_{\bs H_{k,j}}$) densities are bounded so that $\zeta^{-1} \le f_{\bs H\kk} (\bs h) \le \zeta$ and $\zeta^{-1} \le f_{\bs H_{k,j}}(\bs h_j) \le \zeta$ for all $\bs h\in \mc H\kk$ and $\bs h_j\in \mc H_{k,j}$ for some constant $\zeta \ge 1$ and  $j=1,2,...,d\kk$ except that there may exist a finite number of $\bs h_j\in \mc H_{k,j}$ along each coordinate $j$.
\end{assumption}

Assumption~\ref{a:covariate} allows non-hyperrectangular or categorical history spaces. For example, the baseline survival times, $(B_{1} \le B_{2} \le B_{3} \le ...)$, have hypertriangular-shaped support, and the treatment history could be categorical.

\begin{assumption}[Terminal stage sample size]
\label{a:sampleSize}
There exists a constant $\CNK\in(0,1)$ and an integer $\nNK$ such that, for all $n\ge \nNK$, $n_K \ge \CNK n$ almost surely.
\end{assumption}

Assumption \ref{a:sampleSize} enables the use of large sample theory for all intermediate and terminal stages. One such example in the dynamic treatment regime setting is the existence of constants $M_1, M_2\in(0,1)$ and subsets $ \mc N_k\subset\mc H_k$ such that $\Pr(H\kk \in \mc N_k) < M_1, \Pr(U\kk < T\kk \wedge C\kk\mid \bs h\kk) = 0$ for all $\bs h\kk\in \mc N_k$ but
$\Pr(U\kk < T\kk \wedge C\kk \mid \bs h\kk) > M_2$
for all $h\kk\not\in \mc N_k$ for all $k<K$.

In the following theorem, we establish a uniform consistency of the generalized random survival forests, which are constructed using the estimates of the next-stage survival probability ($\hat S_{k+1}$).
This theorem considers a setting where the current stage treatment ($A_k$) could be a part of history ($H_k$), and no optimization over treatment is yet made. 
The proof is provided in Section \ref{as:uc} of the Supplementary Material.


\begin{theorem}[Uniform consistency of a generalized random survival forest]
\label{t:uc}
Suppose Assumptions \ref{a:terminalnode}--\ref{a:sampleSize} hold.
Let $\hat S_{k+1}$ be an estimator of the next-stage survival probability $S_{k+1}$ that is uniformly consistent in both $\bs h\in \mc H_k$ and $t\in [0,\tau]$, 
for stage $k$. Then, the generalized random survival forest $\hat S_k(t\mid \bs h)$ built based on $\Big\{ \big(H_{k,i}, \delta_{k,i}, \hat S_{k+1}(\cdot - X_k $ $\mid H_{k+1,i})\big) \Big\}_{i=1}^{n_k}$ is uniformly consistent. Specifically, for each $k = 1,2,...,K$,
$$\sup_{t\in[0,\tau], \bs h_k\in\mc H_k}|\hat S_k(t\mid \bs h_k) - S_k(t\mid \bs h_k)|\to 0,$$ in probability as $n\to\infty.$
\end{theorem}

We now provide non-asymptotic tail bounds for $\hat S_{k, (b)}$, or the stage $k$ random survival forest estimator constructed using a single random draw per subject $\{(L_{k,i,(b)},\delta_{k,i}): L_{k,i,(b)} \sim \hat S_{k+1}(t-X_k|\bs H_{k+1}), i=1,...,n_k\}$.
We leave it as future work to derive the uniform rate of convergence result for the generalized random survival forests ($\hat S_k(t|\bs h;  S_{k+1})$), or random forests based on multiple or infinitely many draws per subject. However, since the asymptotic variance of $\hat S_k$ can be shown to be smaller than or equal to that of $\hat S_{k,(b)}$, $\hat S_k$ is believed to have at least the same rate of uniform convergence.

To derive a uniform rate of convergence for the survival curves, we assume a different set of assumptions than those used in Theorem \ref{t:uc}. 
We will have a stratified random split assumption---which will be defined soon and is a stronger condition than Assumption \ref{a:split}---and a terminal node size assumption that is more relaxed than Assumption \ref{a:terminalnode}. We also assume a stronger assumption on the covariate distribution---continuous history spaces on unit hypercubes---which could be refined, but we leave it as future work.

\begin{definition}
\label{def:stratified}
A tree is called a `stratified random split tree,' if for each terminal node made of $s$ splits, the number of splits on each variable is greater than or equal to $\lfloor \frac s d \varphi \rfloor$, where $d$ is the number of covariates, $\varphi > 0$ is fixed, and $\lfloor a\rfloor$ is the greatest integer less than or equal to $a$.
\end{definition}

One way to achieve the stratified random splitting is to make splits only among the candidate variables that have not attained the required budget, or $\lfloor \frac s d \varphi \rfloor$ splits, at every other $r$ intermediate nodes, where $r\ge 1$.

\begin{assumption}[Stratified random split trees]
\label{a:stratify}
Trees are constructed under stratified random splitting with a fixed constant $0 < \varphi < 1$.
\end{assumption}

\begin{assumption}[Terminal node size 2]
\label{a:terminalnode2}
The minimum size $n_{\text{min}}$ of the terminal nodes grows satisfying the following rate:
$\lim_{n\to\infty} \frac{\log n \log\log n}{n_{\text{min}}} = 0$.
\end{assumption}

\begin{assumption}[Weakly dependent historical information at each stage 2]
\label{a:covariate2}
The patient history information $\bs H\kk$ at stage $k$ is given as a $d\kk$-dimensional vector lying in $\mc H\kk=[0,1]^{d\kk}$
with its density bounded so that $\zeta^{-1} \le f_{\bs H\kk} (\bs h) \le \zeta$ for all $\bs h\in \mc H\kk$ for some constant $\zeta \ge 1$.
\end{assumption}

Now we state Theorem \ref{t:rateNA}. The proof is provided in the Supplementary Material.

\begin{theorem}[Non-asymptotic bounds for $\hat S_{k,(b)}$]
\label{t:rateNA}
Suppose Assumptions \ref{a:lipschitz2} and \ref{a:covariate2} hold. Let a single-draw-per-person random survival forest $\hat S_{k,(b)}$ be built under Assumptions \ref{a:split}, \ref{a:stratify}, and \ref{a:terminalnode2} for each stage $k=K,K-1, ...,1$. Then, for each $k$, there exists an $n_0$  such that for all $n > n_0$ the following holds with probability $\ge 1 - \frac {2(K-k+1)} {\sqrt {n}}$:
\begin{align}
\label{eq:rate.bound3}
 \sup_{t\le \tau, \bs h_{k}} |\hat S_{k, (b)}(t; \bs h_{k}) - S_{k}(t; \bs h_{k})|  \le r_n,
\end{align}
where $r_n$ is precisely defined in Section \ref{as:rateNA} of the Supplementary Material and has the same rate as $r'_n = \max\Bigg\{ \sqrt{\frac{ \log (n) \{\log(n_{\min}) + \log\log(n) \}   }{n_{\min}}}, \left(\frac{n_{\min}}{ n}\right)^{\frac{\log((1-\alpha)^{-1})}{\log(\alpha^{-1})}\frac{0.991\varphi}{\max_l d_l}}\Bigg\}$.
\end{theorem}




\subsection{Convergence properties of the dynamic treatment regime estimator}
\label{ss:consist2}
We give convergence properties of the dynamic treatment regime estimator using the causal inference framework. With this framework, the theoretical properties are applicable to broader settings than sequentially randomized experiments. A counterfactual outcome is defined as an outcome that would have been obtained if a person had a treatment option contrary to his/her actual treatment. To denote the counterfactual outcomes pertaining to a counterfactual treatment or treatment policy, we add a superscript to the corresponding random variables or survival functions. For example, $S^{\bs\pi}\kk(t|\bs H\kk)$ is a counterfactual survival probability of the remaining life if treatment was given to a person with history $\bs H\kk$ according to a treatment rule $\bs\pi$ from stage $k$ to $K$.
We assume the following three standard causal inference assumptions \citep{hernan2020, rubin2005, cole2009}. 
\begin{assumption}[Stable unit treatment value assumption, or SUTVA]
\label{a:sutva}
Each person's counterfactual dynamics, such as failure time, time to next treatment, and censoring, are not affected by the treatments or history of other patients. Moreover, each treatment has only one version, or if there are multiple versions, their differences are irrelevant. 
\end{assumption}

\begin{assumption}[Sequential ignorability]
\label{a:ignorable}
Treatment assignment is given independently of the counterfactual outcomes, conditionally on the individual's historical information, where the outcomes include all future random quantities such as failure time, time to next treatment, and censoring time. 
\end{assumption}

\begin{assumption}[Positivity]
\label{a:positivity}
For each stage $k=1,2,...,K$, given historical information, the probability of having each treatment $a\in \mc A\kk$ is greater than a constant $L> 0$. Or, $\Pr(A\kk = a|\bs H\kk = \bs h) >L $ for all $a\in\mc A\kk$, $\bs h$ such that $\Pr(\bs H\kk = \bs h) > 0.$
\end{assumption}



The following theorem states that the dynamic treatment regime estimator has a value consistent for the value of the optimal regime. We define the value $\V$ of a treatment regime $\bs\pi$ as the criterion value of the survival probability if all individuals in a population follow the treatment regime. We use $\V(\bs\pi) = \phi(S^{\bs\pi})$ to denote the value of $\bs \pi$, and $\V$ inherits the superscript of $\phi$.

\begin{theorem}[Consistency of the dynamic treatment regime estimator]
\label{t:regime}
Let $\hat{\bs \pi}$ denote the optimal dynamic treatment regime estimator that is built following Algorithm \ref{al:dtrSurv} and Assumptions \ref{a:terminalnode}--\ref{a:split}. Assume that, for each stage, the generalized random forests are built either at once with $\mc A_k\subset \mc H_k$ (``pooled forests") or for each treatment arm (``separate forests"). Further, assume  Assumptions \ref{a:lipschitz2}--\ref{a:sampleSize},  and \ref{a:sutva}--\ref{a:positivity}. 
Then, for each $\phi = \phi^\mu, \phi^{\sigma, t_0}$,
$$|\V(\bs \pi) - \V(\hat{\bs\pi})|\to 0,$$ in probability as $n\to\infty.$
\end{theorem}

Now we present Theorem \ref{t:regimeRate}, a rate of convergence result. Proofs of both theorems and a subsequent corollary are given in Section \ref{as:regime} of the Supplementary Material.

\begin{theorem}[Rate of convergence of the dynamic treatment regime estimator]
\label{t:regimeRate}

Suppose Assumptions \ref{a:lipschitz2}, \ref{a:covariate2}, and \ref{a:sutva}--\ref{a:positivity} hold. Let the single-draw-per-subject random survival forest $\hat S_{k,(b)}$ in Theorem \ref{t:uc} now be built under Assumptions \ref{a:split},\ref{a:stratify}--\ref{a:terminalnode2} for each treatment $a\in\mc A_k$ and each stage $k=1,2...,K$ to build $\hat{\bs \pi}$. 
Then, for each $\phi = \phi^\mu, \phi^{\sigma, t_0}$, 
the following is true in probability as $n\to\infty$: 
\begin{align}
\V(\bs \pi) - \V(\hat{\bs\pi})
& = \mc O_P(\sqrt {r_n}).\nonumber
\end{align}
\end{theorem}

\begin{corollary}[Polynomial rate of convergence]
\label{t:regimeRate2}
Assume the same conditions as in Theorem \ref{t:regimeRate}. Then, $n_{\min}$ can be chosen so that, for some $\eta > 0$,
\begin{align}
\label{eq:rate.poly}
\V(\bs \pi) - \V(\hat{\bs\pi}) = \mc O_P(n^{-\eta/2}).
\end{align}
\end{corollary}


\section{Simulations}
\label{s:sim}

We perform simulations to study the performance of the proposed method and compare those results to that of existing methods. The overall scheme of the simulations follows the pattern in \citet{goldberg2012}, where they mimicked a cancer clinical trial using the tumor size and wellness dynamics. We adapt their data generating mechanism so that dependence among censoring, tumor size, and wellness is incorporated in the mechanism, and deterministic processes are relaxed into more realistic ones.

In the hypothetical trial, patients can go through up to $K=3$ treatment rounds during the trial with length of the trial, $\tau = 10$. Each patient has tumor size $\rho(t)$ and wellness $\omega(t)$ at time $t\in[0,10]$ that affect the timing of failure or treatment. For each stage, patients are given either a more aggressive ($A=1$) or a less aggressive ($A=0$) treatment.

The patient failure time and censoring time are governed by the respective hazard processes:
\begin{align}
    \label{eq:sim_failure}
    \lambda_F(c) & = \frac 1 5\Bigg\{\frac{\rho(u)}{\omega(u)}e^{\bs g(H\kk)^\top \bs \beta_F(A\kk)}  + 1\Big(\omega(u) < 0.1\Big)\Bigg\},\\
    \lambda_C(u) &= e^{\bs g(H\kk)^\top \bs \beta_C(A\kk)},\nonumber
\end{align}
where the tumor size ($\rho$) and wellness ($\omega$) follow the Ornstein-Uhlenbeck process, $g(H\kk) = \{1, \log(B\kk + 1), {\bs Z\kk}^\top\}^\top$, 
$\bs \beta_F(a)$ and $\bs \beta_C(a)$ are the hazard ratios for the failure and censoring time given treatment $a$ (Table~\ref{tab:scenario}), and
the baseline covariates at the $k$th stage follow a $p-$variate normal distribution conditioning on the previous stage value. 
The treatment is allocated according to a Bernoulli trial with propensity 
\begin{align}
\label{eq:propensity}
\pi(\bs H\kk) = \Big\{1 + \exp(-g(\bs H\kk)^\top \bs \beta_\pi)\Big\}^{-1},
\end{align}
where 
the coefficient is $\bs \beta_\pi = (-1, 1, -\frac 1 2 \bs 1_p^\top)^\top$ for the observational data setting and $\bs \beta_\pi = \bs 0_{p+2}$ for the randomized trial setting. See Section \ref{as:simulation} of the Supplementary Material for a detailed description of the dynamics.

\begin{table}
    \def~{\hphantom{0}}
	{\scriptsize
        \begin{tabular}{ccccc}
          \hline
          & Scenario 1 & Scenario 2 & Scenario 3 & Scenario 4\\
          & (base)  & (small $p$) & (large $p$) & (moderate censoring)\\
          \hline
          $p$ & 5& 2& 10 &5\\
          $\bs\beta_F(1)$ & 
          $(0, -3, 2,2,1, -1,-1)^\top$& 
          $(0, -2, 2, 0)^\top$&
          $(0, -2, \bs 2_4^\top,- \bs 1_4^\top, -\bs 2_2^\top)$&
          $(0, -3, 2,2,1, -1,-1)^\top$\\
          $\bs\beta_F(0)$ & 
          $(0, -1, 1,1,1, 1,1)^\top$&
          $(0, -1, 1, 1)^\top$&
          $(0, -1, \bs 1_4^\top,\bs 1_4^\top, \bs 0_2^\top)$&
          $(0, -1, 1,1,1, 1,1)^\top$\\
          $\bs\beta_C(1)$ & 
          $(-3, .2, \bs{.2}_3^\top, -\bs{.2}_2^\top)^\top$&
          $(-3, .2, .2, -.2)^\top$&
          $(-3, .2, \bs{.2}_4^\top, -\bs{.2}_3^\top, \bs 0_3^\top)^\top$&
          $(-2, .2, \bs{.2}_3^\top, -\bs{.2}_2^\top)^\top$\\
          $\bs\beta_C(0)$ & 
          $(-3, .1, \bs{.2}_3^\top,\bs{.2}_2^\top)^\top$&
          $(-3, .1, .2, .2)^\top$&
          $(-3, .1, \bs{.2}_4^\top,\bs{.2}_3^\top, \bs 0_3^\top)^\top$&
          $(-2, .1, \bs{.2}_3^\top,\bs{.2}_2^\top)^\top$\\
          \hline
        \end{tabular}
       }
    \label{tab:scenario}
    \caption{The failure time and censoring hazard coefficients and dimensions of covariates in Scenarios 1 to 4.
        $\bs\beta_F(\cdot)$ and $\bs\beta_C(\cdot)$, the log hazard ratios of the counterfactual failure and censoring times, respectively.}
\end{table}

Simulations are run with two sample sizes, $n = 300$ and $1000$. For each simulation setting, we have $n_{rep} = 200$ simulation replicates, and we estimate the optimal treatment rule based on both the $\phi^\mu$ and $\phi^{\sigma,t_0=5,\mu}$ criteria. We compare the proposed method with the methods in \citet{goldberg2012} and \citet{simoneau2019}. We further compare the results with the observed policy characterized by (\ref{eq:propensity}) and the estimated optimal zero-order model, where all patients are given the same embedded treatment regime, which is the best on average over all embedded regimes. In particular, in implementing the \citet{goldberg2012} method, the linear Q-function space and the random forest-based approximation space were separately used. 
The values of the estimated regimes are evaluated by generating an uncensored sample of size $n_{eval} = 10000$ according to the estimated policies.

\begin{figure}
\includegraphics[width = \textwidth]{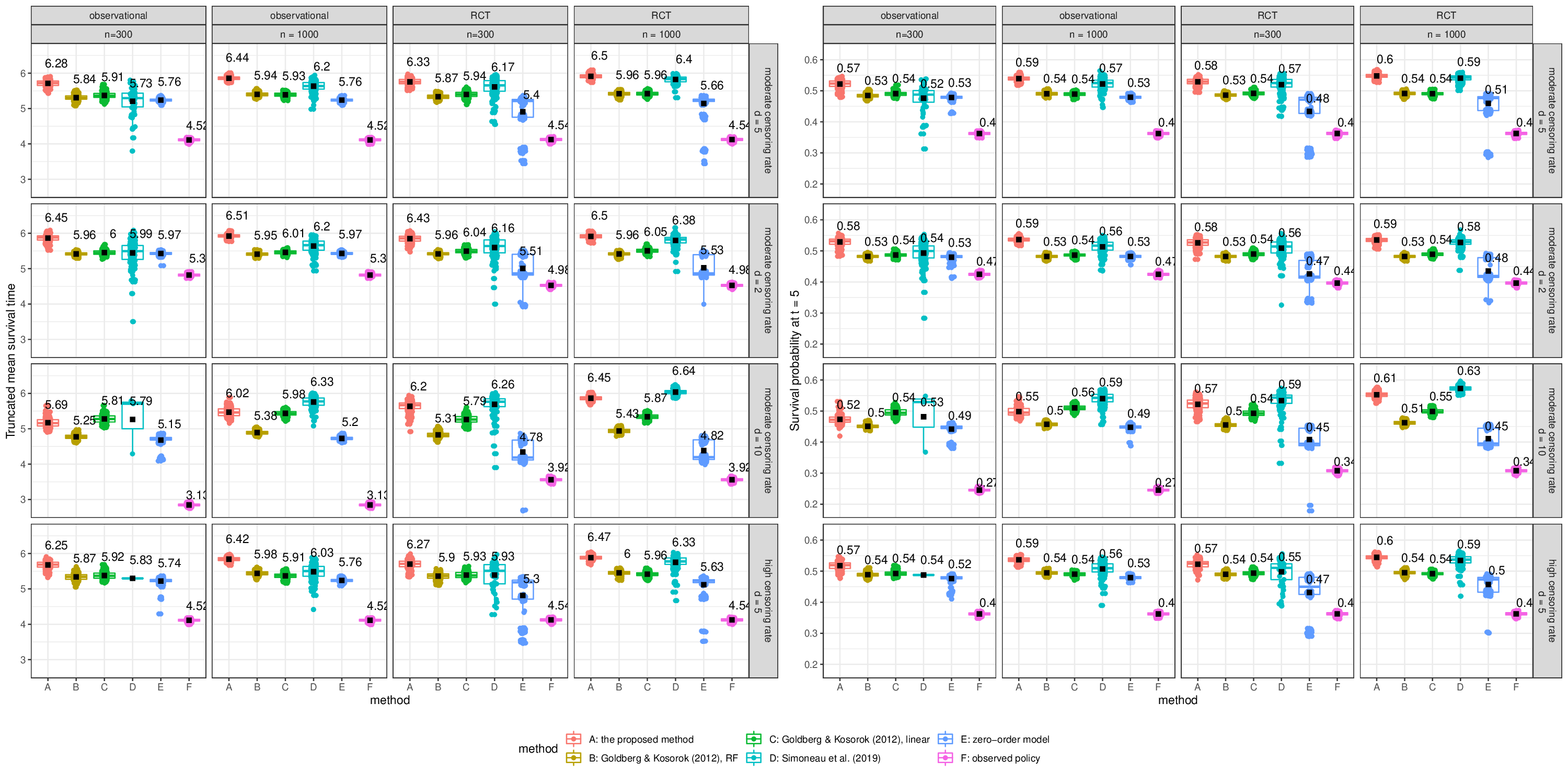}
\caption{The estimated mean truncated survival time (LEFT) and the estimated survival probability at time $t=5$ (RIGHT) under each estimated policy. 
Each dot represents one simulation replicate. The box plots are the quartile summary of the points and the black dots are the average values of each method. Each row of boxes represents a data generation scenarios, and each column of boxes represents a combination of a study design and a sample size.}
\label{fig:sim1}
\end{figure}

Figure~\ref{fig:sim1} LEFT shows how the truncated mean survival time ($\phi^\mu$) of each estimated policy behaves under different settings. 
For most of the settings, on average, all estimated regimes have greater values than the zero-order model, and the zero-order model has a higher value than the observed policy. This implies that the standard of care improves the overall survival time of patients over the observed policies and that individualized treatment rules can further enhance the outcomes.

Figure~\ref{fig:sim1} LEFT also shows that among the individualized treatment rules, the proposed method outperforms the other methods in many settings. Even when the method exhibits lower or about the same performance as the other methods for the small sample size, it often has higher performance than the others for the larger sample size. See, for example, the high censoring cases (the last row) of either the observational or the RCT setting. In the settings where the dimension of the covariate is high (the third row), the method of \citet{simoneau2019} often has high values.
This might be because of the doubly robustness property of the method and the fact that the true optimal decision rule might be close to a linear function, as suggested by the linear interaction of the stage-wide hazard functions in (\ref{eq:sim_failure}).
The method of \citet{simoneau2019}, however, often does not provide estimates when either the sample size of the terminal stage is small due to censoring or failure or the dimension of covariates is large.

Figure~\ref{fig:sim1} RIGHT shows the simulation results in terms of the survival probability at $t_0=5$, or $\phi^{\sigma, 5}$.
Note that the estimated regimes for the proposed method are distinct between LEFT and RIGHT but are the same for the other methods.
The proposed estimator shows better performance in terms of the policy values under most settings, giving us similar interpretations as the previous results.

The computational complexity for estimating a dynamic treatment regime is $\mc O(n^2\log(n) $ $\sum_{k=1}^K|\mc A_k||\mc H_k|)$. Using the \texttt{R} package \texttt{dtrSurv}, the estimation of a three-stage regime in these simulations takes less than several (20) seconds for $n=300$ ($n=1000$).

\section{Data examples---The Atherosclerosis Risk in Communities (ARIC) Study}
\label{s:examples}
We illustrate the use of dynamic treatment regimes using two biomedical studies---the Atherosclerosis Risk in Communities (ARIC) Study and a study of acute myeloid leukemia. The latter analysis is provided in Section \ref{as:leukemia} of the Supplementary Material.

We derive an optimal dynamic treatment regime for cardiovascular failure events using the ARIC Study. ARIC is a community-based observational cohort study where a total of 15,792 individuals aged 45--64 had been recruited at four centers in U.S. (Washington County, MD; Forsyth County, NC; Jackson, MS; and Minneapolis, MN) in 1987--1989 \citep{wright2021}. The data are available on approval of the ARIC Data Coordinating Center (\url{https://sites.cscc.unc.edu/aric/distribution-agreements}).
The main goal of ARIC is to identify cardiovascular risk factors. Each participant was examined for their medical, social, behavioral, and demographic information up to seven times until 2019. We analyze the most recent data with enough number of survival events---the fifth visit ($n=6527$, in 2011--2013) and the sixth visit ($n=3996$, in 2016--2017). The seventh visit ($n=3853$, in 2018--2019) was not included due to a small number of observed events. 

Out of the 6,527 subjects who made at least five visits, we further narrowed down the focus to those who already had or had had heart diseases or heart failure ($n=945$) at their baseline visit. The age range of this high-risk group is 66--90 years at the beginning of their fifth visit, and among them,  69\% were White, 53\% were female, 15.1\% and 11.4\% had experienced prevalent heart failure and coronary heart diseases at their first visit, respectively, and 41\% died of circulatory diseases 
until $\tau=2700$ days from the fifth visit. We regard deaths due to other causes as censored. Administrative censoring includes those who survived 2700 days. Of those who made visits 5 and 6, 31\% and 57\% were censored, respectively. Besides the cause-specific mortality, we also derive all-cause mortality-based dynamic treatment regimes---i.e., other cause deaths are also regarded as failure---and the resulting censoring rates are 12.6\% and 50\% for the two visits.

The ARIC study followed participant use of Statin (HMG CoA reductase inhibitors) and anti-coagulants, which have been shown to be effective in reducing cardiovascular diseases and mortality \citep{mills2008, johansen2014, schneeweiss2012}.
These medications, however, may not be as effective and can even be detrimental for some patients. For example, long-term use of anti-coagulants could increase the risk of cardiovascular calcifications \citep{schurgers2004}.
Statin, although it has been considered safe for most patients, still has side effects such as type II diabetes and has interactions with multiple medications \citep{thompson2016, ramkumar2016}. Thus, a personalized dynamic strategy for prescribing these medications could enhance long-term survival outcomes, especially for those who had experienced heart disease or failure. 

We develop two-stage optimal dynamic treatment regimes for prescribing a Statin and anti-coagulant combination. We follow the 80-20 cross-validation scheme for the value function estimation described in Section~\ref{as:aric} of the Supplementary Material. The numbers of patients in each treatment arm are 279 (none),  492 (Statin),  60 (anti-coagulant), and 114 (both) at visit 5, and the corresponding numbers for visit 6 are 110, 230, 38, and 68. 
The tailoring variables include gender, race, age, number of years on cigarette, the  baseline prevalent heart failure history, the baseline coronary heart disease history, and heart failure history at visit 5. Also, for each of visits 5 and 6, BMI, waist-to-hip ratio, alcohol consumption, hypertension, blood glucose level, smoking status, blood HDL cholesterol level, and Aspirin use are included. For visit 6, the past-visit medication history (i.e., visit 5 medications) is included. We assume that the four-level treatment classification satisfies Assumption \ref{a:sutva}, and also that the effects of the unmeasured confounders are negligible.
Missing values were imputed by the multivariate imputation with chained equations \citep{van2011}. With a small fraction of missing values---less than 3\% for several variables and less than 1\% or none for most---we present the results based on the first imputed dataset. 
Because the \citet{simoneau2019} method was implemented using \texttt{R} package \texttt{DTRreg} the current version (1.7) of which allows binary treatment arms only, the method was applied for finding the optimal anti-coagulant regime instead. Also, as the method suffers from unstable estimation issue with high-dimensional covariates, a subset of the mentioned covariates were considered for the method. More details of the model for the \citet{simoneau2019} method are described in Section \ref{as:aric} of the Supplementary Material.

Figure \ref{fig:aric} illustrates the cross-validation-based value estimates of each estimator.
The proposed method provides the largest average survival time and six-year (2200 days) survival probability than other methods for both optimization criteria ($\phi^\mu$, $\phi^{\sigma, 2200}$) and event types (cardiovascular, all-cause). The zero-order model recommends Statin for visit 5 and both for visit 6 for most of the cross-validation replicates. On the other hand, the proposed method has various prescription options for different subjects. The higher values of the suggested method over the zero-order model values highlight the benefit of individualized treatment rules. The higher performance of the suggested method over the other competing methods is likely attributed to the model misspecification issues in other methods, such as linearity and independent censoring.

\begin{figure}
    \centering
    \includegraphics[width = 0.9\textwidth]{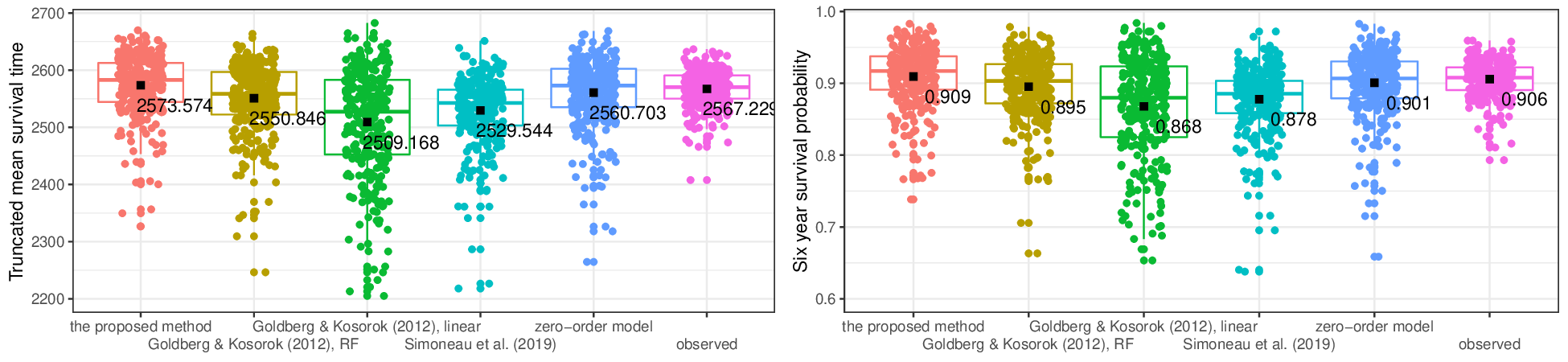}
        \includegraphics[width = 0.9\textwidth]{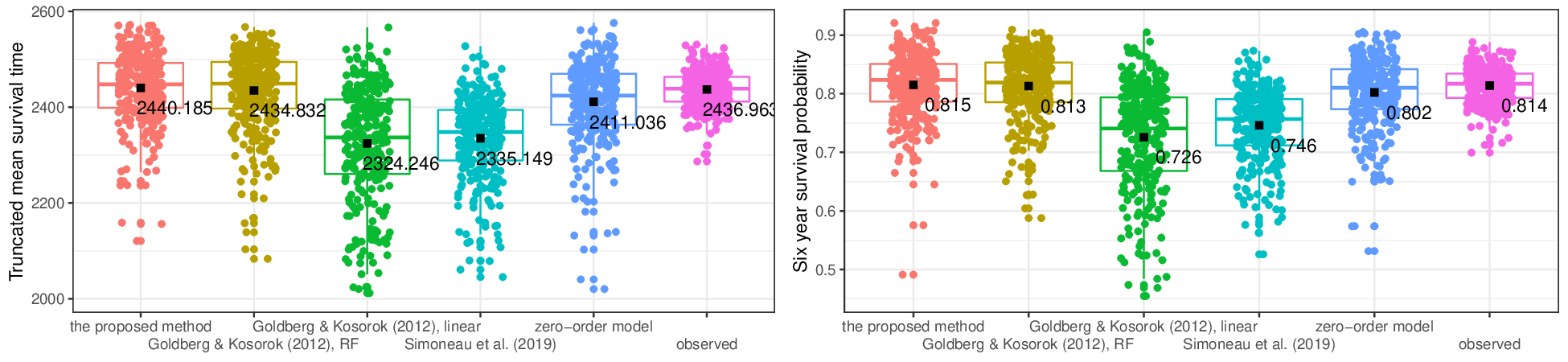}
    \caption{Estimated truncated mean survival time (LEFT) and estimated six-year survival probability (RIGHT) of the dynamic treatment regime estimators for preventing the cardiovascular-specific (TOP) and all-cause (BOTTOM)  mortality. Each dot represents one cross-validation replicate. The black square dots are the average values of each method.}
    \label{fig:aric}
\end{figure}

\section{Discussion}
\label{s:discussion}
In this paper, we introduced a new dynamic treatment regime estimator for survival outcomes. The estimator is versatile in the sense that it allows for a flexible number of treatment stages and arms, it gives a non-linear decision rule that maximizes either the mean survival time or the survival probability, and it permits dependent censoring. In this section, we discuss some considerations when using our method and how the method can be further extended.

The proposed estimator assumes that the distribution is well characterized by the treatment stages rather than the natural passage of time. That is, the cohort at each treatment stage might include patients that received the stage treatment on day 10 as well as patients that received the stage treatment on day 1000. Thus, this method is effective when the disease dynamic is considered stationary between stage transitions or when the treatment effect contains a relatively stronger signal than that of the baseline disease dynamics. 


The proposed method assumes an unrestricted policy class. In practice, however, clinicians and patients may prefer understandable, simple rules. A linear rule, for example, is often of interest. By posing a Cox-type proportional hazards assumption on top of our method, the resulting policy class becomes a set of linear classifiers. The task then reduces to replacing the generalized random survival forest with the generalized Cox model, which essentially does a Cox regression based on multiple random draws.

In this paper, treatment timing ($U_{k}$) was  not given as a policy variable but as a part of history for the next stage. Recently, \cite{nie2021} worked on finding optimal treatment regimes that provide the best treatment timing. Finding a timing for optimizing the overall survival probability is interesting future work.

Another interesting extension of the proposed method is optimizing a specific quantile of the survival distribution, such as the median survival time.
However, this task introduces a unique challenge with respect to the backward recursion. Specifically, the optimal decisions made for later stages may not be optimal once earlier decisions moderate the distribution. Thus, an exhaustive search may be needed to find an optimal policy. An extension of the quantile-optimal dynamic treatment regime estimators developed by \citet{wang2018} and \citet{linn2017} to right-censored data would be interesting future work.


\section*{Acknowledgement}

This research was funded in part by grant P01 CA142538 from the National Cancer Institute.
The authors thank Donglin Zeng for bringing the composite criterion optimization into the discussion and the editors and anonymous reviewers for their constructive feedback. The ARIC study has been funded in whole or in part with Federal
funds from the National Heart, Lung, and Blood Institute, National Institutes of Health,
Department of Health and Human Services, under Contract nos. (HHSN268201700001I,
HHSN268201700002I, HHSN268201700003I, HHSN268201700004I, HHSN268201700005I). The
authors thank the staff and participants of the ARIC study for their important contributions. 
The authors also thank Peter F. Thall, Abdus S. Wahed, and Yanxun Xu for providing the leukemia study data.

\section*{Supplementary material}
\label{SM}
Supplementary Material available online includes 
the summary of notation in Section \ref{as:notation}, more details about the generalized random survival forests in Section \ref{as:grsfmethod}, the proofs of the theorems in Sections \ref{as:prop1}--\ref{as:regime}, and more details about the simulations and data examples in Sections \ref{as:simulation}--\ref{as:dataExamples}.

\bibliographystyle{biometrika}
\bibliography{dtrSurv_R2}

\end{document}